\documentclass[prd,twocolumn,a4paper,superscriptaddress]{revtex4}
\usepackage{graphicx,amssymb}
\begin{document}
\newcommand{\be}{\begin{equation}}
\newcommand{\ee}{\end{equation}}
\newcommand{\bq}{\begin{eqnarray}}
\newcommand{\eq}{\end{eqnarray}}
\newcommand{\bsq}{\begin{subequations}}
\newcommand{\esq}{\end{subequations}}
\newcommand{\bc}{\begin{center}}
\newcommand{\ec}{\end{center}}
\newcommand {\R}{{\mathcal R}}
\newcommand{\al}{\alpha}
\newcommand\lsim{\mathrel{\rlap{\lower4pt\hbox{\hskip1pt$\sim$}}
    \raise1pt\hbox{$<$}}}
\newcommand\gsim{\mathrel{\rlap{\lower4pt\hbox{\hskip1pt$\sim$}}
    \raise1pt\hbox{$>$}}}

\title{Linear and non-Linear Instabilities in Unified Dark Energy Models}
\author{P. P. Avelino}
\email[Electronic address: ]{ppavelin@fc.up.pt}
\affiliation{Centro de F\'{\i}sica do Porto, Rua do Campo Alegre 687, 4169-007 Porto, 
Portugal}
\affiliation{Departamento de F\'{\i}sica da Faculdade de Ci\^encias
da Universidade do Porto, Rua do Campo Alegre 687, 4169-007 Porto, 
Portugal}
\author{L.M.G. Be{\c c}a}
\email{luis.beca@fc.up.pt}
\affiliation{Centro de F\'{\i}sica do Porto, Rua do Campo Alegre 687, 4169-007 Porto, 
Portugal}
\affiliation{Departamento de F\'{\i}sica da Faculdade de Ci\^encias
da Universidade do Porto, Rua do Campo Alegre 687, 4169-007 Porto, 
Portugal}
\author{C.J.A.P. Martins}
\email[Electronic address: ]{C.J.A.P.Martins@damtp.cam.ac.uk}
\affiliation{Centro de F\'{\i}sica do Porto, Rua do Campo Alegre 687, 4169-007 Porto, 
Portugal}
\affiliation{Centro de Astrof\'{\i}sica, Universidade do Porto, Rua das Estrelas s/n, 4150-762 Porto,
Portugal}
\affiliation{Department of Applied Mathematics and Theoretical 
Physics,
Centre for Mathematical Sciences,\\ University of Cambridge,
Wilberforce Road, Cambridge CB3 0WA, United Kingdom}

\date{24 Janeiro 2008}
\begin{abstract}
We revisit the paradigm of unified dark energy discussing in detail the averaging problem in this type of scenarios, highlighting the need for a full non-linear treatment. We also address the question of if and how models with one or several dark fluids can be observationally distinguished. Simpler and physically clearer derivations of some key results, most notably on the relation between the generalized Chaplygin gas and the standard ($\Lambda$CDM) `concordance' model and on a Jeans-type small-scale instability of some coupled dark energy/dark matter models are presented.
\end{abstract}
\maketitle

\section{\label{intr}Introduction}

Observations have been steadily piling over the years that can only be `explained', in the context of General Relativity, by the presence of so-called \emph{dark} forms of energy: ordinary baryonic matter and radiation are simply not enough \cite{Riess,Perlmutter,Percival,Dodelson,Spergel}. In broad terms, a dark energy component (violating the strong energy condition) is required to explain the recent acceleration of the Universe and a (cold) dark matter component to account for the amount of structure observed.  

Unfortunately, these components have never been observed directly. In fact, they may not \emph{exist} at all. It is certainly possible to engineer dark energy and dark matter out of the picture, by modifying General Relativity directly; often this process involves extra-dimensions with non-trivial topologies and/or fiddling with TeVeS gravity sources \cite{Teves} (see also \cite{Freese,Alternatives}). At the moment, however, most modifications seem far too \emph{ad hoc} to be particularly pleasing. In this paper, we will mostly work within the confines of General Relativity, i.e., we will assume that dark energy and dark matter are real entities. On the other hand, if this assumption is correct, it is truly mind boggling to find out that roughly $96\% $ of the Universe should be in this dark form. Cosmology is thus hard-pressed to explain what these components are. 

Certainly the realization that dark energy and dark matter may not be \emph{independent} entities should play a significant role. Although we frequently perceive them as being different, this does not have to be the case. The Chaplygin gas \cite{Chap1,Chap2}, for instance, experienced a short burst of popularity precisely for  being able to mimic both dark energy and dark matter (depending on the \emph{local} density), and still be just one form of energy---an exotic one, granted, but the same can be said of quintessence, $k$-essence, and the rest of the phenomenological zoo. Thus, we should be open to the possibility that dark energy and dark matter are actually just  different manifestations of the same underlying field, not different entities \emph{per se}.  On the theoretical front, the idea of a unified description of dark energy and dark matter has incredible heuristic potential and should be properly investigated. In case of failure, we would at least come back justified in treating dark energy and dark matter as an independent pair. Either way, something of value could be gained.

While the background and linear properties of unified dark energy models have been extensively studied by the community, the non-linear aspects of the problem have been largely ignored. In the following section, we will discuss in detail the averaging problem in the context of these models and  
show that non-linearities are a crucial aspect  that cannot be neglected.

\section{Averaging in unified dark energy models}

The real Universe (henceforth denoted by manifold $\mathcal M$) displays hierarchical structures like stars, galaxies, clusters of galaxies, and so on; it is far from being smooth. The dynamics in $\mathcal M$ is assumed to be entirely described by General Relativity (or some modified version of), as gravity is  the relevant force at work on cosmological scales. In practice, however, due to the highly non-linear nature of the field equations, it is virtually impossible to solve them fully (even numerically), except for a few high symmetry configurations; see, for instance, \cite{Stephani}.

On the other hand, several observations indicate that the Universe looks increasingly smooth, as  larger and larger scales are considered (typically over $100\, \text{Mpc}$). This \emph{average} background (representing the \emph{global} behaviour of $\mathcal M$) is routinely idealized as a completely featureless manifold, hereafter denoted by manifold $\langle \mathcal M \rangle$. The kinematics in $\langle \mathcal M \rangle$ is essentially contained in the cosmological principle, but what about its dynamics? This is actually a tricky question. Normally, it is assumed that General Relativity applies just as well in $\langle \mathcal M \rangle$ as it does in $\mathcal M$. Yet, when we average $\mathcal M$ to obtain the background, we are also averaging complex non-linear interactions. Averaging linear terms in the field equations is no big deal, but averaging non-linear terms \emph{is}: this is because they introduce backreaction terms and hence new dynamics. 

The common expectation however, is that this backreaction is negligible. Proving this, of course, is very hard; even checking it numerically is extremely difficult. As a matter of fact, `averaging', in General Relativity, remains largely an unresolved problem \cite{Ellis,Barrow}. In connection to this, it has even been suggested that the current acceleration of the Universe might be the result of these corrections, which would avoid the need for an exotic dark energy component (see for example \cite{Buchert} and references therein). As appealing as this may sound, however, such corrections are extremely hard to quantify, and currently there is no convincing argument to show that they are sufficiently large to do the job. The few times that such an analysis has been attempted, by studying high symmetry configurations such as closely spaced sheets of matter separated by voids (which are arguably not a good approximation of the real universe), have shown them to be small, and hence incapable of producing an accelerated behaviour \cite{Gruzinov} (see also \cite{Li}).

A related problem that will be relevant for what follows concerns the averaging of the energy content of the universe described by the energy-momentum tensor of the matter fields. The evolution of $\langle \mathcal M \rangle$ is usually assumed to be described by General Relativity with a source term given by the energy-momentum tensor of a perfect fluid with an energy density $\langle \varepsilon \rangle$, where $\langle \ \rangle$ is an appropriate average, and a known equation of state, $p=p(\varepsilon)$. Even if we take for granted that for $p=0$ backreaction effects are small, we can not ignore the fact that in general $\langle p \rangle \neq p(\langle \varepsilon \rangle)$. An exception to this is the special case where the equation of state is given by $p=w \varepsilon$ where $w$ is a constant; then, we have $\langle p \rangle =w \langle \varepsilon \rangle = p(\langle \varepsilon \rangle)$. However, this is no longer true for the equation of state of the phenomenological model known as the generalized Chaplygin gas \cite{GCG},
\begin{equation}
p=-\frac{A}{\rho^{\alpha}}\,,\label{chapgas}
\end{equation}
where $A$ and $\alpha$ are positive constants. Indeed, it is straightforward to check that
\begin{equation}\label{inequality}
\langle p \rangle  = 
- \langle A / \varepsilon^\alpha  \rangle \neq - A / \langle \varepsilon \rangle^\alpha = 
p(\langle \varepsilon \rangle)\,.
\end{equation} 
This is in fact a crucial problem in the analysis of all unified dark energy models not sufficiently close to a $\Lambda$CDM model (which for the generalized Chaplygin gas is characterized by $\alpha=0$ \cite{LCDM}) for the above effects to be negligible---indeed, this invalidates quite a few simple analyses.  In fact, it has been shown that in some models the transition from matter to dark energy domination may never occur or may do so much earlier than expected in the absence of perturbations \cite{Bassett,NONLIN}. Of course, this problem only occurs when perturbations become non-linear on certain scales. If, on the other hand, perturbations are linear on every scale then
\begin{equation}\label{equality}
\langle p \rangle = \langle p(\langle \varepsilon \rangle (1+ \delta)) \rangle\sim \langle p(\langle \varepsilon \rangle)\rangle\,,
\end{equation}
if $\delta = (\varepsilon-\langle \varepsilon \rangle) / \langle \varepsilon \rangle \ll 1$.

\section{Scalar fields}

In this paper we will use a $(-+++)$ metric signature. Now, let us consider the action
\begin{equation}\label{eq:L}
S=\int d^4x \, \sqrt{-g} \, {\mathcal L}(X,\phi) \, ,
\end{equation}
where $\mathcal L$ is the Lagrangian density, $\phi$ is a real scalar field 
and 
\begin{equation}\label{eq:kinetic_scalar}
X=-\frac{1}{2}\nabla^\mu \phi \nabla_\mu \phi \,,
\end{equation}
is the kinetic term.
As a simple example, consider the case of a classical scalar field $\phi$ governed by the so-called \emph{canonical} Lagrangian
\begin{equation}\label{eq:scalar_L}
{\mathcal L} = X - V(\phi) \, ,
\end{equation}
where $V(\phi)$ is some scalar potential. The energy-momentum tensor may be obtained by varying the action relative to the inverse metric, which yields
\begin{equation}\label{eq:canonical_T}
T_{\mu\nu} (\phi) =  \nabla_\mu \phi \nabla_\nu \phi + (X - V(\phi))g_{\mu\nu} \, .
\end{equation}
If we now make the following identifications
\begin{equation}
u_\mu = \frac{\nabla_\mu \phi}{\sqrt{2X}} \label{eq:velocity} \, , \quad \varepsilon = X+V(\phi) \, , \quad p = X-V(\phi) \, ,
\end{equation}
and plug them into (\ref{eq:canonical_T}) we obtain
\begin{equation}\label{eq:fluid}
T^{\mu\nu} = (\varepsilon + p) u^\mu u^\nu + p g^{\mu\nu} \, ;
\end{equation}
hence the energy-momentum tensor of $\phi$ can be written in the form of a perfect fluid. Recall that in (\ref {eq:fluid}), $u^\mu$ is the 4-velocity field describing the motion of the fluid (if $\nabla_\mu \phi$ is timelike) while $\varepsilon$ and $p$, are its proper energy density and pressure, respectively.

The above is easily generalized for an arbitrary Lagrangian density of the form ${\mathcal L}(X,\phi)$ \cite{Mukhanov}. This defines a new class of scalars broad enough for our purposes. Again, we can still explicitly rewrite the energy momentum tensor for this model in a perfect fluid form, by means of the following identifications
\begin{equation}\label{eq:new_identifications}
u_\mu = \frac{\nabla_\mu \phi}{\sqrt{2X}} \, ,  \qquad \varepsilon = 2 X p_{,X} - p \, , \quad p =  {\mathcal L}(X,\phi)\, .
\end{equation}
From this it follows that if $p=p(X)$, then $\varepsilon = \varepsilon(X)$. Unfortunately, it is not always possible to invert $\varepsilon(X)$, and  obtain $X(\varepsilon)$ but when it is, the fluid has an \emph{explicit} isentropic equation of state $p=p(\varepsilon)$. A useful example is $p \propto X^n$ which, as one can easily check, describes a constant $\omega = 1/(2n-1)$ fluid. In particular, when $n=0$  the scalar corresponds to a standard cosmological constant, $n=1$, to a massless scalar field, $n=2$, to background radiation, and so on. In the limit of large $n$, the scalar field can be interpreted as dust (a pressureless non-relativistic fluid).

How about the generalized Chaplygin gas? We can retro-engineer (\ref{eq:new_identifications}) to ascertain the necessary Lagrangian by solving
\begin{equation}
\varepsilon = 2X \alpha \, \frac{A}{\varepsilon^\alpha}\, \frac{\varepsilon_{,X}}{\varepsilon} + \frac{A} {\varepsilon^\alpha} \, ,
\end{equation}
a non-linear differential equation. Dividing by $\varepsilon$ and rewriting everything in terms of a new function $\xi=A/\varepsilon^{1+\alpha}$, we end up with a much nicer linear version
\begin{equation}
1 = -\frac{\alpha}{1+\alpha} 2X \xi_{,X} + \xi \, ,
\end{equation}
which is easy to solve. The solution is 
\begin{equation}
\xi=1-(2X)^{\frac{1+\alpha}{2\alpha}}
\end{equation}
and thus the Lagrangian which reproduces the generalized Chaplygin becomes
\begin{equation}\label{eq:L_gCg}
{\mathcal L}(X)=p(X) = -\frac{A}{\varepsilon^\alpha} = -A^\frac{1}{1+\alpha} 
\xi^\frac{\alpha}{1+\alpha}\, ,
\end{equation} 
with $0 <2 X < 1$.

Finally, the sound speed for any scalar field (canonic or otherwise) is also well defined and determined in linear theory to be \cite{Mukhanov}
\begin{equation}
c_s^2 \equiv \frac{p_{,X}}{\varepsilon_{,X}}=\frac{{\mathcal L}_{,X}}{{\mathcal L}_{,X}+2X{\mathcal L}_{,XX}}\,;
\end{equation}
contrast this to the isentropic result $\delta p / \delta \varepsilon$. It follows that quintessence has a constant sound speed of unity. This is another reason why quintessence cannot reproduce the generalized Chaplygin gas. On the other hand, if we apply this formula to the scalar field governed by (\ref{eq:L_gCg}), the sound speed does come out the same as the isentropic one and all is well.

\section{Interlude: The $\Lambda$CDM limit}

Implementing the generalized Chaplygin gas as a scalar field obeying a particular Lagrangian is useful on a number of levels. For one, it allows us to prove that the $\alpha=0$ limit of the generalized Chaplygin gas is totally equivalent to an ordinary $\Lambda$CDM model. This fact plays an important role when comparing unified dark energy models with observations. While obviously true in the absence of perturbations, the need to explicitly demonstrate this beyond a zero-order equivalence, only became apparent after \cite{Fabris} appeared. In it the surprising claim was made that the linear evolution of perturbations for $\Lambda$CDM and the $\alpha=0$ generalized Chaplygin gas model actually differed. Nevertheless, in \cite{LCDM} it was shown that this was not true (see also \cite{Quercellini}; their equivalence to first order in the metric perturbations was established and it was equally argued that the correspondence went well beyond linear order. Here we give a much simpler proof of this equivalence. If we expand (\ref{eq:L_gCg}) in a power series and take the limiting case $\alpha \rightarrow 0$ we obtain
\begin{eqnarray}
p(X) &=& \lim_{\alpha \to 0} -A^{1/1+\alpha} \left[1 - \frac{\alpha}{1+\alpha} (2X)^{\frac{1+\alpha}{2\alpha}} \right]   \nonumber \, ,\\
 &=& -A + 0 \cdot (2X)^\infty = -A\, .
\end{eqnarray}
Thus, everywhere in $\mathcal M$, the Lagrangian decomposes nicely into a cosmological constant plus `matter', thus demonstrating the equivalence between $\alpha=0$ and $\Lambda$CDM to any order; gravity alone does not distinguish between the two. Moreover, given that both are, to a certain extent, simply toy-models with somewhat shaky motivations from fundamental physics, this is probably as far as they can be meaningfully compared.

\section{A truly `atomic' fluid?}

We have shown that the $\alpha=0$ limit of the generalized Chaplygin gas is equivalent to $\Lambda$CDM. An obvious follow-up question is whether this equivalence between a single unified dark energy fluid and some family of minimally coupled components is valid in general. Recently \cite{Kunz} has argued that it is always possible to split a single unified dark energy fluid into different minimally coupled components or to combine several fluids into a single fluid behaving in exactly the same way as the original mixture, from a cosmological point of view. Although this is obvious in the absence of perturbations, it was also argued that this degeneracy went beyond the background level.

Before trying to answer the above question, we need to clearly define what we mean by a single fluid. It is clear that by allowing the complexity of the fluid to be arbitrarily large, for example by considering very high order tensor fields, we may in principle get disproportionately non-trivial dynamics. However, in this case, do we have one fluid or several minimally coupled fluids? In general, a complex single fluid description is also prone to a multi-fluid interpretation. In this paper, we shall refer to a single indivisible (`atomic') fluid as one whose Lagrangian density is a function of a real scalar field which has the form $\mathcal L(X,\phi)$. We shall focus our attention on the isentropic subfamily characterized by an equation of state of the form $p=p(\varepsilon)$.

It goes without saying that the energy-momentum tensor of any fluid can always be split into several pieces. In fact there is an infinite number of ways to accomplish this. The critical question, however, is how to interpret any such decomposition. Are the resulting parts real \emph{physical} fluids? Do they exist \emph{independently} from one other? We have already seen that the answer is \emph{no} if the generalized Chaplygin gas does originate from the single scalar field governed by (\ref{eq:L_gCg}). Each piece is then a \emph{virtual} component, i.e., without independent existence (except in the special $\alpha=0$ case). In fact, the evolution of individual \emph{virtual} parts is not, in general, constrained by causality. Conversely, if we are in the presence of various fluids we can also add their energy-momentum tensors. However, the dynamics of the resulting fluid could be very complex and in general it will not be describable by a real scalar field with a Lagrangian density of the form $\mathcal L(X,\phi)$.

From a cosmological point of view all the relevant information is contained in the energy momentum tensor which acts as a source for the gravitational field. Pending laboratory evidence (which in principle \emph{can} detect not only fields themselves but even couplings between them) we are only sensitive to the total energy-momentum tensor and consequently \cite{Kunz} argues that cosmology alone does not provide useful information on whether a single unified dark energy fluid or a family of minimally coupled fluids is responsible for the observations.

However, if we have a single fluid described by an arbitrary equation of state $p=p(\varepsilon)$ and consider the evolution of very large wavelength perturbations in a homogeneous and isotropic universe we know that, given local initial conditions for $H$ and $\varepsilon$, the subsequent evolution is the same in any such patch. This is no longer true if we consider a family of minimally coupled fluids. Let us consider large wavelength flat patches characterized by the same value of $H$ (flatness implies that the value of $\varepsilon$ is also the same in any such patches). It is then easy to show that we may find any number of families of three minimally coupled fluids with the same initial conditions for $\varepsilon$, ${\dot \varepsilon}$, $p$, and ${\dot p}$ but with very different subsequent evolutions. As an example consider the two families of three fluids characterized by a constant $w$ with
\begin{itemize}
\item $w_1=-1=-w_3$ and $w_2=0$, with $\varepsilon_i/\varepsilon=1/3$ for $i=1,2,3$
\item $w_1=-{\sqrt 6}/3=-w_3$, with $\varepsilon_j/\varepsilon=1/2$ for $j=1,3$ and $\varepsilon_2/\varepsilon=0$
\end{itemize}
We see that if we are in the presence of various fluids, a similar behaviour at a given time may well lead to very different behaviours at a later time. Moreover, we are even allowed to decide how much freedom we wish to give ourselves: we may set up adiabatic or iso-curvature fluctuations (or indeed a combination of them). Clearly, the former situation will be more restrictive than the latter.

\section{Linear instabilities in unified dark energy models \label{sound}}

The simplest unified dark energy model (the generalized Chaplygin gas in the $\alpha=0$ limit) has a vanishing sound speed $c_s^2=0$ but this is no longer the case in the context of more general models. It is easy to show \cite{Beca} that the evolution of the sound speed in the absence of non-linear perturbations is fully determined by the background evolution and is equal to
\begin{equation}
\label{eq5}
c_s^2 \equiv \frac{d p}{d \rho}=\frac{1}{3{\rm{H}}}
\frac{d}{d{\rm{H}}}\left[{\rm{H}}^2\left(q-\frac{1}{2}\right)\right] \, ,
\end{equation}
where $q \equiv -\ddot a / (a {\rm{H}}^2)$ is the usual deceleration parameter. The sign of $c_s^2$ is thus linked to the background dynamics and in particular to how fast the transition from dark matter to dark energy occurs. 

A simple illustration of the above point is provided by the generalized Chaplygin gas equation of state given by Eqn. (\ref{chapgas}), which corresponds to
\begin{equation}\label{eq:omega_gCg}
w = -A/ \varepsilon^{\, 1+\alpha} \, ;
\end{equation}
phenomenologically we allow $\alpha$ to take any value. For $\alpha > 0$ this is the generalized Chaplygin gas itself, for $\alpha=0$ this is equivalent to $\Lambda$CDM and for $\alpha=-1$ and $A \le 1$ it is equivalent (in the absence of perturbations only) to a standard quintessence model with constant $w$. We shall not consider the case with $\alpha < -1$ since this would imply a universe which is dark energy dominated at early times and we are only interested in the late time behaviour of dark energy. This fluid illustrates the above points. For $\alpha > -1$ its energy density evolves as
\begin{equation}\label{eq:gCg_energy}
\varepsilon  = \varepsilon _0 \left[ {\cal A} +  (1 - {\cal A})  \left(\frac{a_0}{a} \right)^{3(1 + \alpha)}  \right]^{1/1 + \alpha},
\end{equation}
where ${\cal A} = A/\varepsilon_0^{1+\alpha}$ is a constant. The sound speed has the form $c_s^2=-\alpha w$ and consequently $c_s^2 > 0$ for $\alpha > 0$. We clearly see that the sign is related to the speed of the transition from the matter dominated era to the dark energy dominated era. If it is steep enough (specifically, faster than in $\Lambda$CDM), $c_s^2$ will be positive, otherwise it will be negative.  

On the other hand, consider a universe with matter and a standard scalar field with the Lagrangian of Eqn. (\ref{eq:scalar_L}) and consider what happens if we 
treat cold dark matter (CDM) and dark energy (DE) as a single isentropic fluid. This may be justifiable, to a certain extent, if there is a strong coupling between the dark matter and dark energy sectors as is the case in \cite{Kaplinghat,Bean,Bean2}. In the absence of perturbations, the energy density of this effective fluid satisfies
\begin{equation}
\label{eq:dot_rho_over_rho}
\frac{\dot {\varepsilon}}{\varepsilon} = - 3 H (1+\omega) \, ,
\end{equation}
with $w=p_{\rm DE}/(\varepsilon_{\rm DE}+\varepsilon_{\rm CDM}) > -1$. Since $\varepsilon > 0$ we obviously have $\dot \varepsilon < 0$, and the sign of $c_s^2$ 
only depends on the pressure evolution which is given by
\begin{equation}
\label{eq:dot_p_over_t}
\frac{dp}{dt} = 
{\dot \phi}\left(\ddot \phi - \frac{dV}{d\phi}\right) \sim -{\dot \phi} \frac{dV}{d\phi} = - {\dot V} > 0 \, ,
\end{equation}
where the slow-roll approximation has been used. Hence, in this case 
\begin{equation}
\label{eq:sound_speed_2}
c_s^2 = \frac{\dot p}{\dot \varepsilon} < 0\, .
\end{equation}
This instability \cite{Beca} will occur in any case where $\dot p > 0$ (assuming $\dot \varepsilon < 0$). A particular example of this is the Jeans-type instability studied in \cite{Afshordi,Bjaelde,Kaplinghat,Bean,Bean2}, which occurs for models where there is a sufficiently strong coupling between dark matter and dark energy. (Note, however, that such large couplings are strongly constrained by several equivalence principle type experiments, so the cosmological relevance of these models remains unclear.) If $w_{\rm DE}$ is a constant, it is straightforward to show that
\begin{equation}
\label{eq:sound_speed_3}
c_s^2 = \frac{w_{\rm DE}(1+w_{\rm DE})\varepsilon_{\rm DE}}{(1+w_{\rm DE})\varepsilon_{\rm DE}+\varepsilon_{\rm CDM}} \, .
\end{equation}
Interestingly, the models where this instability is absent are precisely those where one can get $w_{\rm DE}<-1$ (as long as $w_{\rm DE} > -1 - \varepsilon_{\rm CDM}/\varepsilon_{\rm DE}$) \cite{Beca,Afshordi,Kaplinghat,Bean,Bean2}. Hence, the above argument for  $c_s^2 < 0$ no longer holds if one allows $w_{\rm DE}<-1$.

\section{\label{conc}Conclusions}

In the so-called concordance cosmology model, a range of observational data 
is used to postulate the existence of two dark fluids for which so far there 
is no direct experimental evidence. It is therefore natural to consider 
scenarios where these fluids are coupled, or even different manifestations 
of  a single fluid. 

In this paper, we have re-visited unified dark energy models and 
discussed some of their distinguishing features. In particular, we have 
addressed in some detail the averaging problem, showing that non-linear
effects cannot, in general, be ignored. We have also clarified 
the relation between a particular class of UDE models and the standard 
$\Lambda$CDM paradigm, and the physical reason behind a Jeans-type instability 
that has been identified in sufficiently coupled models.


We have equally discussed the question of whether models with one or several dark fluids can be observationally distinguished showing that in the latter case, similar initial conditions for the total energy density and pressure can lead to very different outcomes. This degeneracy  is absent in single fluid models.

Finally, we would like to point out that a laboratory detection of one or more dark 
components is not the only  way of ultimately 
breaking this degeneracy. Equivalence principle tests, for example, can 
provide key information on the dark sector. Even more promising is the prospect 
of using astrophysical measurements of varying couplings to reconstruct both 
the dark energy equation of state and a measure of the dark sector's coupling 
to the standard model as a function of redshift \cite{alpha}. This will be further explored 
in future work.


\bibliography{UDE}

\end{document}